\documentstyle[psfig]{mn}

\begin{document}
\title[Primeval Galaxies in HDF-S]{Candidate High Redshift and
Primeval Galaxies in Hubble Deep Field South}
\author[D.L. Clements, S.A. Eales, A.C. Baker]
{D.L. Clements$^{1}$, S.A. Eales$^{1}$, A.C. Baker$^1$\\
$^1$Department of Physics and Astronomy, University of Wales Cardiff,
PO Box 913, Cardiff, CF2 3YB\\ }

\maketitle

\begin{abstract}
We present the results of colour selection of candidate high redshift
galaxies in Hubble Deep Field South (HDF-S) using the Lyman dropout
scheme. The HDF-S data we discuss were taken in a number of different
filters extending from the near--UV (F300W) to the infrared (F222M) in
two different fields. This allows us to select candidates with
redshifts from z$\sim$3 to z$\sim$12. We find 15 candidate z$\sim$3
objects (F300W dropouts), 1 candidate z$\sim$4 object (F450W dropout)
and 16 candidate z$\sim$5 objects (F606W dropouts) in the $\sim$ 4.7
arcmin$^2$ WFPC-2 field, 4 candidate z$\sim$6 (optical dropouts) and 1
candidate z$\sim$8 (F110W dropout) in the 0.84 arcmin$^2$ NICMOS-3
field.  No F160W dropouts are found (z$\sim$12). We compare our
selection technique with existing data for HDF-North and discuss
alternative interpretations of the objects. We conclude that there are
a number of lower redshift interlopers in the selections, including
one previously identified object \cite{t98}, and reject those objects
most likely to be foreground contaminants. Even after this we
conclude that the F606W dropout list is likely to still contain
substantial foreground contamination.  The lack of candidate very high
redshift UV-luminous galaxies supports earlier conclusions by Lanzetta
et al. \shortcite{l98}.  We discuss the morphologies and luminosity
functions of the high redshift objects, and their cosmological
implications.

\end{abstract}


\section{Introduction}

Recent years have seen significant advances in our understanding of
supposedly normal galaxies at high redshift (z$>$2). This has been
made possible by the introduction of simple photometric techniques
that permit the successful selection of faint high redshift
galaxies. These techniques include full-scale photometric redshift
estimation methods (eg. Fernandez-Soto et al. \shortcite{f98}), but
perhaps the simplest is the Lyman-dropout
method. This was pioneered by Steidel et al. \shortcite{s95}, and
relies on the passage of the ubiquitous 912\AA~Lyman-limit
discontinuity through broad-band imaging filters.  At higher redshifts
the suppression of light between 1216\AA~ and 912\AA~ in the emitted
frame by the Lyman-$\alpha$ forest becomes significant, so that
1216\AA~ becomes the break-point in the spectra \cite{sp98}.  This
approach was initially used with purpose built U,G and R filters
\cite{s95} to select galaxies at z$\sim$3. The presence of the 912\AA~
discontinuity in the U band leads to a large U-G colour while the flat
spectral energy distribution redward of the Lyman break produces small G-R
colours. The use of imaging filters allows large areas to be surveyed,
with the candidate high redshift galaxies later re-observed for
spectroscopic confirmation. Several surveys have been conducted using
this approach, and we are now able to probe the large scale
distribution of z$\sim$3 star-forming galaxies (eg. Steidel et
al. \shortcite{s98a}) and determine luminosity functions and colour
distributions (eg. Dickinson et
al. \shortcite{di98}). It is also possible to extend this approach to
higher redshifts by looking for the Lyman break at redder
wavelengths (eg. Steidel et al. \shortcite{s98c}).

Much of the work on high redshift galaxies in the original Hubble Deep
Field (HDF-N) used the Lyman-break method to identify candidate
objects (eg. Clements \& Couch \shortcite{c96}, Madau et
al. \shortcite{m96}). The deep, high resolution imaging provided by
HST allows selection of good candidates with lower luminosities than
the ground-based Lyman-break galaxies, and permits studies of
morphology that were previously very difficult. Application of the
method redward of the bluest HDF-N filter, F300W, has also permitted
the selection of candidate galaxies at z$\gg$3, two of which have been
spectroscopically confirmed at z=5.6 \cite{w98} and 5.34 
(Spinrad et al., \shortcite{sp98}) and several
others at 4$<$z$<$5.

The main disadvantage of HDF-N, is its very small area on the
sky. This means that the volume surveyed is a very small
`pencil-beam'.  When combined with the observed clustering of these
objects \cite{s98a}, there may be significant problems in drawing any
statistical conclusions (eg. Madau et al. \shortcite{m96}) solely from
HDF-N galaxies.

The HDF-South observations (HDF-S) thus offer us a significant
potential increase in our knowledge of the high redshift
universe. Apart from doubling the area surveyed with the Wide Field
and Planetary Camera 2 (WFPC-2), the addition of the NICMOS infrared
and STIS optical/UV observations allows us to significantly extend the
redshift limits of the dropout technique. In principle, bright
galaxies at z$>$12 might even be detected as dropouts at 1.6$\mu$m,
appearing only in the longest wavelength band, 2.2$\mu$m. The present
paper applies simple Lyman-dropout selection techniques to the WFPC2
and NICMOS HDF-S data to provide initial lists of candidate galaxies
out to the highest redshifts obtainable for use in further studies and
in spectroscopic followup. We also discuss the implications of these
results on our understanding of galaxy formation and the star
formation history of the universe.

We use AB magnitudes throughout this paper and assume H$_0$ =
50kms$^{-1}$Mpc$^{-1}$, q$_0 = 0.5$ and $\Lambda = 0$.

\section{Selection of High Redshift Candidates}

The HDF-S data discussed here consists of observations from two
separate regions. Firstly, the WFPC2 data, consisting of deep images
in the F300W, F450W, F606W and F814W (3000\AA, 4500\AA, 6060\AA~and
8140\AA~respectively) filters reaching 10$\sigma$ magnitude limits of
26.8, 27.7, 28.2 and 27.7 (in a 0.2 sq. arcsec. area)
respectively \cite{hdfs} and covers an area of 4.7 sq. arcmin..
Secondly, the NICMOS Camera 3 region observed with the F110W, F160W,
F222M filters (1.1$\mu$m, 1.6$\mu$m and 2.2$\mu$m respectively) and
the STIS open CCD (STIS50) channel (covering 2000\AA~to 1.1$\mu$m)
reaches 10$\sigma$ magnitude limits of 27.0, 26.8, 24.0 and 28.4
(in 0.8 sq. arcsec. area) respectively and covers 0.84 arcmin$^2$
\cite{hdfs}.  Using the Lyman-break method we can thus, in principle,
search for galaxies to redshifts $>$12 using the F222M band.

To apply photometric selection criteria we must first extract
photometric source catalogues for HDF-S. To achieve this we apply the
SExtractor programme \cite{b96} to the combined drizzled images of the
HDF-S provided by STScI \cite{hdfs}. For WFPC2, objects are
selected initially in the reddest filter, F814W, and then photometric
parameters in other filters are measured at the positions of these
objects. The objects are selected with a detection threshold of
1.5$\sigma$ above the background and a minimum area of 0.4 arcseconds
(ie. 10 connected pixels). This produces a total of 2257 detected
objects. We use the ISO\_MAG parameter extracted by SExtractor for
source magnitudes, which gives the isophotal magnitude for an object
down to 1.5$\sigma$ above the background. This parameter is stable at
faint limits and avoids any uncertainties resulting from correction to
total magnitudes, though such corrections should amount to only
$\sim$5\%.  This process produces a catalogue including magnitudes in
all four passbands for each object detected in the F814W images.

For NICMOS, a somewhat different procedure is applied.  Firstly, the
STIS CCD image, which has a final pixel scale three times smaller than
the drizzled NICMOS images, must be rebinned to match the NICMOS pixel
scale to facilitate catalogue matching.  Then, when searching for
objects missing from filter $a$ we select objects in filter $b$, the
next reddest wavelength. Thus a catalog of objects missing from the
STIS image, but detected at F110W, is based on a source list derived
from the F110W filter, and similarly F110W dropouts are derived from a
source list based on F160W detections. The SExtractor detection
parameters are similar to those used for WFPC2. The F110W selected
list contains 292 sources, the F160W list 342, and the F222M list 51,
because of the lower magnitude limit due to the higher background in
this filter.

Once the catalogues have been generated we select candidate high
redshift objects using a technique derived from the extensive followup
observations of HDf-N. Consideration of the colour-colour diagram for
HDF-N galaxies in Dickinson \shortcite{di98} shows that the following
criteria will select the majority of high redshift objects
in a given filter, $a$ if:
\begin{enumerate}
\item The colour between the filter $a$ and the next reddest filter $b$ $>$1.5
\item The object should have been detected with at least
10$\sigma$ significance in filter $a$ if the colour $a - b = 1.5$
\item There is no more than 1 magnitude difference between the flux in adjacent
filters redder than $b$
\end{enumerate}
Criterion 2 provides the magnitude limits for the present selection of
candidate high z galaxies ie. F450W$<$25.3 for F300W dropouts,
F606W$<$26.2 for F450W dropouts, F814W$<$26.7 for F606W dropouts,
F110W$<$26.9 for STIS dropouts, F160W$<$25.5 for F110W dropouts, and
F222M$<$24 for any F160W dropouts.

With these criteria we would select 9 of the 12 F300W dropout galaxies
above our magnitude limit with spectroscopically confirmed redshifts
in HDF-N \cite{di98}. We would also select all of the candidate z$>$4
objects selected by photometric redshift methods above our magnitude
limit, of which two have confirmed redshifts, given by Fernandez-Soto
et al. \shortcite{f98}. Lower redshift interlopers are a potential
problem.  We can assess their contribution by examining the number of
interlopers we would have selected in HDF-N. In Fernandez-Soto's
photometric and spectroscopic redshift compendium, we would select 17
F300W dropouts of which only 7 have either a spectroscopic or
photometric redshift below 2. For F450 dropouts we would select just
two objects, both at z$>$3.4, while for F606 dropouts we would select
27 objects, of which only 5 lie at z$>$ 4.5. 
Comparable data is not
available in the infrared to test the NICMOS selections.

\section{Results}

Using the above colour and magnitude criteria, we search our
SExtractor catalogues for candidate high redshift galaxies from
z$\sim$3 to z$\sim$12.  The images of these candidate objects are then
examined, leading to the rejection of several objects lying on the
edge of, or beyond, the areas surveyed in one or more filters. The
final catalogue of candidate high redshift objects is listed in Table
1. We find 15 possible F300W dropouts, candidate z$\sim$2.5--3.5
objects, one z$\sim$3.5--4.5 F450W dropout, 16 candidate F606W
z$\sim$4.5--5.5 dropouts, four candidate z$\sim$2.3--7.0 STIS dropouts
(though one, and possibly two, are lower redshift very red objects -
see below).  We find no candidate F110W or F160W dropouts within our
magnitude limits, though we do find one fainter F110W dropout
candidate which would lie at z$\sim$5.6--9.7. Images of the F300
dropout objects are
shown in Figure 1.

\begin{table*}
\begin{tabular}{ccccccc} \hline
WFPC2 Cat. No.&RA(J2000)&Dec(J2000)&F814&F606&F450&F300\\ \hline
\multicolumn{7}{c}{z$\sim$ 3 F300 Dropout Candidates} \\ \hline
\\
  38&22 32 53.98&-60 34 20.53&24.76$\pm$0.08&24.96$\pm$0.04&25.08$\pm$0.13&26.60$\pm$0.6\\
 149$^1$&22 32 53.33&-60 34 13.69&22.83$\pm$0.02&23.21$\pm$0.01&23.88$\pm$0.04&26.14$\pm$0.26\\
 257&22 32 59.91&-60 34 05.02&24.14$\pm$0.02&24.90$\pm$0.02&25.29$\pm$0.06&27.01$\pm$0.47\\
 760$^1$&22 32 58.10&-60 33 37.08&23.13$\pm$0.01&23.60$\pm$0.01&24.29$\pm$0.03&26.28$\pm$0.29\\
 877&22 33 04.89&-60 33 29.48&24.31$\pm$0.03&24.43$\pm$0.01&24.68$\pm$0.04&$>$27.8\\
 880&22 32 50.64&-60 33 28.84&24.45$\pm$0.03&24.72$\pm$0.02&25.07$\pm$0.05&$>$27.8\\
 895&22 33 03.19&-60 33 28.77&23.34$\pm$0.02&23.68$\pm$0.01&24.02$\pm$0.03&26.20$\pm$0.25\\
1158&22 33 03.87&-60 33 12.89&24.56$\pm$0.04&24.77$\pm$0.02&25.15$\pm$0.07&$>$26.6\\
1562&22 32 57.22&-60 32 41.50&24.55$\pm$0.04&24.78$\pm$0.02&25.03$\pm$0.07&27.00$\pm$0.69\\
1745&22 32 49.00&-60 32 26.85&23.44$\pm$0.02&23.61$\pm$0.01&24.14$\pm$0.03&$>$27.8\\
1748&22 32 49.22&-60 32 27.09&23.00$\pm$0.01&23.20$\pm$0.01&23.68$\pm$0.03&$>$26.9\\
1816&22 32 47.62&-60 32 20.61&24.43$\pm$0.03&24.84$\pm$0.02&25.26$\pm$0.06&$>$26.9\\
1847$^1$&22 32 47.76&-60 32 17.59&22.72$\pm$0.01&23.15$\pm$0.01&23.72$\pm$0.03&25.71$\pm$0.16\\
1934$^1$&22 32 46.51&-60 32 08.38&22.97$\pm$0.02&23.40$\pm$0.01&24.07$\pm$0.04&26.61$\pm$0.41\\
1951&22 32 48.87&-60 32 06.71&24.11$\pm$0.02&24.32$\pm$0.01&24.54$\pm$0.04&26.07$\pm$0.21\\
\\ \hline
\multicolumn{7}{c}{z$\sim$ 4 F450 Dropout Candidates} \\ \hline
\\
1153&22 32 52.56&-60 33 15.47&24.70$\pm$0.03&25.62$\pm$0.02&27.27$\pm$0.17&$>$27.8\\
\\ \hline
\multicolumn{7}{c}{z$\sim$ 5 F606 Dropout Candidates} \\ \hline
\\
  17&22 32 59.66&-60 34 20.06&24.12$\pm$0.03&25.75$\pm$0.04&26.77$\pm$0.23&$>$26.25\\
  50$^1$&22 33 00.58&-60 34 17.08&23.47$\pm$0.02&25.11$\pm$0.02&26.90$\pm$0.21&$>$27.5\\
 211$^1$&22 32 57.79&-60 34 08.40&23.03$\pm$0.01&25.29$\pm$0.02&27.07$\pm$0.16&$>$27.8\\
 494$^1$&22 32 55.39&-60 33 55.01&23.30$\pm$0.02&25.45$\pm$0.04&27.03$\pm$0.24&$>$26.6\\
 601&22 33 05.11&-60 33 45.00&26.00$\pm$0.06&27.96$\pm$0.09&$>$28.7&$>$27.5\\
 618&22 32 58.63&-60 33 46.22&25.83$\pm$0.05&27.51$\pm$0.07&$>$28.7&$>$27.8\\
 836&22 33 00.00&-60 33 33.45&24.98$\pm$0.03&26.68$\pm$0.04&27.94$\pm$0.26&$>$27.8\\
 872&22 32 56.80&-60 33 31.50&25.37$\pm$0.04&26.95$\pm$0.05&28.01$\pm$0.27&$>$27.3\\
 899&22 32 57.07&-60 33 28.73&24.24$\pm$0.02&25.94$\pm$0.03&27.35$\pm$0.21&$>$27.8\\
1270&22 32 51.60&-60 33 06.55&24.38$\pm$0.03&26.14$\pm$0.04&27.67$\pm$0.26&$>$27.8\\
1364$^1$&22 32 57.98&-60 32 58.71&24.94$\pm$0.06&27.50$\pm$0.10&$>$28.7&$>$27.8\\
1473&22 32 57.96&-60 32 49.56&25.01$\pm$0.05&26.54$\pm$0.06&27.65$\pm$0.35&$>$26.5\\
1757&22 32 49.63&-60 32 27.64&26.58$\pm$0.08&28.40$\pm$0.10&$>$28.7&$>$27.8\\
2005&22 32 53.14&-60 32 01.58&24.81$\pm$0.03&26.66$\pm$0.04&$>$28.7&$>$27.0\\
2051&22 32 49.22&-60 31 58.69&26.35$\pm$0.07&28.00$\pm$0.08&$>$28.7&$>$27.8\\
2068$^1$&22 32 55.94&-60 31 56.49&24.43$\pm$0.03&26.81$\pm$0.05&28.33$\pm$0.35&$>$27.8\\ \hline
NICMOS Cat. No.&RA(J2000)&Dec(J2000)&F222&F160&F110&STIS\\ \hline
\multicolumn{7}{c}{z$\sim$ 6 Optical Dropout Candidates} \\ \hline
\\
  24$^*$&22 32 51.09&-60 39 09.80&21.29$\pm$0.08&21.79$\pm$0.01&22.72$\pm$0.02&25.93$\pm$0.25\\
  99&22 32 55.20&-60 38 52.19&23.9$\pm$0.50&24.16$\pm$0.03&24.32$\pm$0.04&26.57$\pm$0.25\\
 110&22 32 49.10&-60 38 50.75&$>$24.0&27.03$\pm$0.07&26.82$\pm$0.08&29.07$\pm$1.00\\
 255$^2$&22 32 51.22&-60 38 22.56&23.04$\pm$0.16&23.70$\pm$0.01&24.64$\pm$0.02&26.74$\pm$0.25\\
\\ \hline
\multicolumn{7}{c}{z$\sim$ 8 F110 Dropout Candidates} \\ \hline
\\
 189&22 32 48.27&-60 38 43.69&24.9$\pm$0.50&26.36$\pm$0.07&$>$28.0&28.3$\pm$0.70\\
\\ \hline
\multicolumn{7}{c}{z$\sim$ 12 F160 Dropout Candidates} \\ \hline
\\
\multicolumn{7}{c}{None}\\

\end{tabular}
\caption{Details of drop out galaxies}
Limits are 3$\sigma$ upper limits. $^1$ indicates likely low redshift
interloper,
$^*$ Treu et al. (1998) VRO at z$>$1.7.$^2$ indicates other possible low z VRO.
See text for details.
\end{table*}

\begin{figure}
\psfig{file=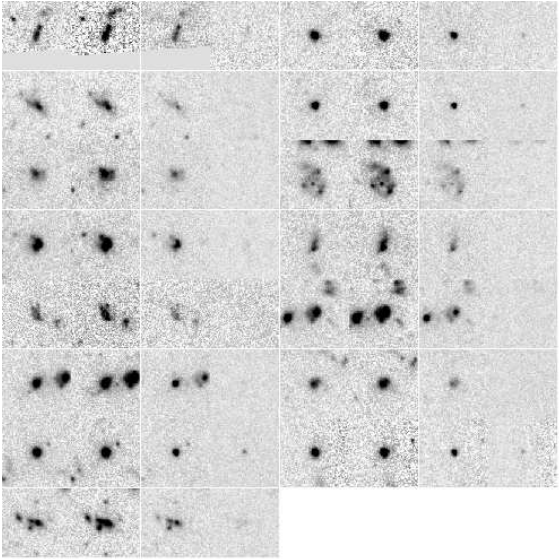,width=8cm}
\label{images}
\caption{Images of F300W dropouts. Images are shown in groups of four horizontally, showing
F814W, F606W, f450W and F300W images from left to right. Objects are shown in the same order
as in Table 1, starting in the top left, then left to right, and then working down the page.
Images are 4 arcseconds on a side. North is up and East is to the right.}
\end{figure}

%\begin{figure}
%\psfig{file=f450dropgal.ps,width=8cm}
%\label{images2}
%\caption{Images of F450W dropout.
%Arranged as in Figre 1.}
%\end{figure}

%\begin{figure*}
%\psfig{file=f606dropgal.ps,width=16cm}
%\label{images3}
%\caption{Images of F606W dropouts.
%Arranged as in Figure 1.}
%\end{figure*}

%\begin{figure}
%\psfig{file=stisdropgal.ps,width=8cm}
%\label{images4}
%\caption{Images of STIS optical dropouts. Images shown are F222M, F160W, F110W, STIS
%from left to right. Images are 7.5 arcseconds across, North is up and East is to the right.}
%\end{figure}

%\begin{figure}
%\psfig{file=f110candim.ps,width=8cm}
%\label{images5}
%\caption{Images of F160W dropout.
%Arranged as in Figure 4.}
%\end{figure}

\section{Discussion}

\subsection{High Redshift, Old or Red?}

The determination of redshift estimates with broad band colours is
a very inexact science, especially when limited to only a
few passbands.  The possibility exists that certain classes of lower
redshift object might be able to mimic the colours of the dropout
galaxies, and thus enter our list of candidate high redshift objects.
Of specific concern are the Balmer decrement at 4000\AA~in old
galaxies, and the recently discovered highly dust obscured galaxies
\cite{c98} known as Very Red Objects \cite{g96}.

The flux step across the 4000\AA~break in ellipticals can be as high
as a factor of 2 to 3 (eg. Coleman et al. \shortcite{c80}). It is thus
possible for old, red elliptical galaxies to masquerade as drop-outs
at certain redshifts. For the F300W dropouts selection is based
on the F300W-F450W colour so low redshift ellipticals might thus appear
as F300W dropouts. Such objects would be bright, since they lie
at lower redshift, and would have the distinctive r$^{1/4}$ surface
brightness profile of an elliptical. Moderate redshift ellipticals
could also interfere with selection in other wavebands.  For
z$\sim$0.2--0.5 they might appear as F450W dropouts, while
at still higher redshift (z$\sim$0.5--1) they could appear as F606W
dropouts. The increasing volume encompassed by the HDFS at higher
redshifts will mean that an increasing number of foreground
ellipticals will lie in the field for higher redshift selections. The
problem of old, red objects is thus likely to be severest for the F606W
dropouts, since there is $\sim$3--4 times as much volume for
foreground interlopers to appear through the 4000\AA~ break than for
F450W dropouts.

The class of objects that has become known as Very Red Objects (VROs)
also pose a problem. VROs are defined by their extremely red colours,
(R-K)$_{vega}>$5, or, equivalently, (R-K)$_{AB}>$3.3.  The nature of
these objects is still unclear. They may simply be old ellipticals at
large redshift, or they may be heavily dust-obscured objects
containing either a starburst or an AGN \cite{e96}. These objects could
clearly appear in the NICMOS selected lists. Indeed, one VRO
originally discussed by Treu et al. \shortcite{t98} which is suggested
to be an elliptical at z$>$1.7, would appear to have been
re-discovered by our selection as object no. 24, the brightest in our
optical dropout selection \cite{t99}.
A further object, no. 255, seems likely to be
a VRO, since it is also bright and significantly extended in the infrared
--- very similar to no. 24.

\subsection{Absence of Very High Redshift Galaxies}

HDF-S observations are capable of finding forming galaxies at very
high redshifts -- up to z$>$12 for F160W dropouts. Galaxies forming
stars at 100 (300)M$_\odot yr^{-1}$ for q$_0 = 0.5 (0.0)$ are
detectable in the F222M filter. However, we find very few candidate
very high redshift galaxies. Only one candidate z$\sim$8 object is
found, and no candidates at higher redshift. This supports the
conclusions of Lanzetta et al. \shortcite{l98} based on optical and
near-IR observations of HDF-N that there are very few UV-luminous very
high redshift galaxies.  Combining the present results with those of
Lanzetta et al., we conclude that the number density of these objects
is $<1.2$ arcmin$^{-2}$.  However, this limit refers only to {\em
unobscured} very high redshift galaxies. Since these observations are
dependent on detecting the far-UV emission from high redshift
galaxies, even a moderate amount of obscuration could dim such objects
below the detection threshold.  The recent detection of a cosmological
infrared background, eg. Puget et al. \shortcite{p96}, Fixsen et al.
\shortcite{fi98}, suggests that obscuration may have a significant role to play
in the high redshift universe. We must thus treat these limits with
caution.

\subsection{Properties of Candidate High z Galaxies}

The sizes of the candidate high redshift objects, as measured by the
second moment of their F814 images for the WFPC2 objects and the F160W
images for the NICMOS objects, range from a barely resolved 0.3'', to
4.0''. It is noticeable that the F300W dropouts are larger, with a
mean size of 1.3''$\pm$0.3'', compared to the more distant galaxies,
with 0.72''$\pm$0.12'' for the F606 dropouts. This does not
necessarily reflect a cosmological effect. As noted above it is
probable that there are more red, foreground elliptical galaxies
contaminating the F606 dropouts. Such objects would have a smaller
size than the often very extended and disturbed genuine high redshift
galaxies. In this context it is interesting to note that the smallest
F300 dropout objects are often the brightest eg. objects 149 and
1847. A similar effect can also be noted in the F606 dropouts. These
objects are also those which most often have some faint emission in
the dropout band, which tends to corroborate the suggestion that they
are lower redshift interlopers. We thus suspect that several of the
smaller brighter objects may be foreground interlopers.  These are
noted in Table 1. A more stringent selection criterion for F606
dropouts using F814 - F606 $>2.0$ instead of 1.5 would select just
three objects in HDF-N, all of which have spectroscopic or photometric
redshifts $>$ 4.5. For HDF-S this selection produces no believable
high redshift candidates. On the basis of this and the number of
interlopers the selection would select in HDF-N, the F606W dropout
list should probably be treated with caution until spectroscopic
redshift confirmations are available.

The morphologies of the candidate high redshift galaxies show an
interesting shift from F300 dropouts to longer wavelength
selections. Many of the F300 dropouts show disturbed extended
morphologies, as is the case for similar objects in HDF-N
\cite{c96}. These include the spectacularly disturbed object 880, and an
apparent interacting pair of dropout galaxies 1745 and 1748, which may
be a similar system to the famous `Hotdog' galaxy in HDF-N \cite{b98}. In
contrast, most of the longer wavelength dropout candidates are smaller
and with much less unusual morphologies. This might indicate that the
higher redshift objects are the original pre-galactic clumps which
then merge together at around z$\sim$3 to form larger objects, or
alternatively might just be due to the cosmological (1+z)$^4$ surface
brightness dimming from z$\sim3$ to z$\sim$5 and above.

\begin{figure}
\psfig{file=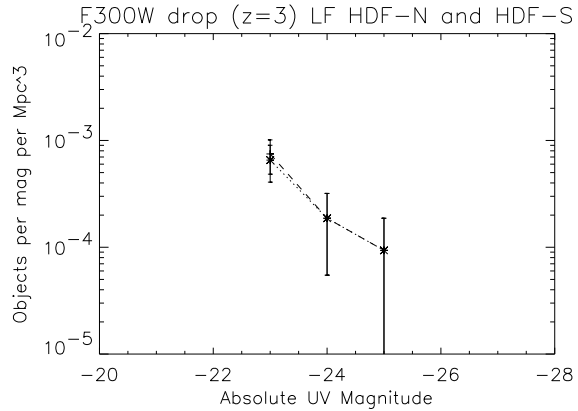,width=8cm}
\caption{z=3 Lyman Dropout Galaxy Luminosity Function}
Comparison of the HDF-N and HDF-S z$\sim$3 Lyman dropout galaxy
luminosity functions. Data for HDF-N from \cite{f98} data for HDF-S
from the present paper, excluding four bright, small objects. HDF-N LF is 
shown by the dotted line, HDF-S uses a dashed line. Note the
good match between the two data sets.
\label{lf}
\end{figure}

Figure \ref{lf} shows a comparison of the UV ($\sim$1600\AA )
luminosity function (LF) of 2.5$<$z$<$3.5 galaxies in Fernandez-Soto
et al. \shortcite{f98}, all of which have spectroscopic confirmation,
with the LF of F300 dropout galaxies from the present paper, after the
removal of the small, bright galaxies discussed above.  As can be seen
the LFs match very well.

\section{Conclusions}

We have selected candidate high redshift galaxies using the
established Lyman `drop-out' technique, but have applied this to the widest
range of wavelengths possible in the HDF-S. We have found candidate
galaxies up to z$\sim$ 8. A number of the dropout selected objects,
though, are identified as low redshift interlopers, demonstrating the
need for spectroscopic followup and observations over a broader
wavelength range.  We note that the z$\sim$3 F300W selected objects
typically show highly disturbed morphologies while candidates at higher
redshifts do not, perhaps indicating that z$\sim$3 saw the onset of
substantial galaxy or galaxy-subunit merging.  We will next develop
more sophisticated photometric redshift modelling and will apply this
to broader wavelength observations of HDF-S as they become available.

{\bf Acknowledgements} It is a pleasure to thank the HDF and HDF-S
teams at STScI and STECF and E. Bertin for SExtractor. The anonymous referee
made many useful comments that have improved the paper. DLC and ACB are
supported by PPARC postdoc grants.

\end{document}